\documentclass[letterpaper,english,reprint, aps]{revtex4-1}
\usepackage[T1]{fontenc}
\usepackage{booktabs}
\usepackage{amsmath}
\usepackage{amssymb}
\usepackage{graphicx}
\usepackage{subscript}

\usepackage{xcolor}

\makeatletter

\pdfpageheight\paperheight
\pdfpagewidth\paperwidth

\providecommand{\tabularnewline}{\\}

\makeatother

\usepackage{babel}

\begin{document}
\title{Third-order polarizability of interlayer excitons in hetero-bilayers}
\author{M. F. C. Martins Quintela$^{1,2}$, J. C. G. Henriques$^{1,2}$, N. M. R. Peres$^{1,2}$}
\address{$^{1}$Department and Centre of Physics, University
of Minho, Campus of Gualtar, 4710-057, Braga, Portugal}
\address{$^{2}$International Iberian Nanotechnology Laboratory (INL), Av. Mestre
Jos{\'e} Veiga, 4715-330, Braga, Portugal}
\begin{abstract}
In this paper, we employ a fully microscopic approach to the study of interlayer excitons in hetero-bilayers. We use Fowler's and Karplus' method to access the dynamical polarizability of non--interacting interlayer excitons in a $\mathrm{WSe}_{2}/\mathrm{WS}_{2}$--based van der Waals heterostructure. 
Following from the calculation of the linear polarizability, we consider Svendsen's variational method to the calculation of the dynamic third--order polarizability. 
With this variational method, we study both two--photon absorption and third--harmonic generation processes for interlayer excitons in a $\mathrm{WSe}_{2}/\mathrm{WS}_{2}$ hetero--bilayer, discussing the various selection rules of intra--excitonic energy level transitions.
\end{abstract}
\maketitle

\section{Introduction}

The advent of the study of two--dimensional layers with atomic thickness
has renewed the interest in transition--metal dichalcogenides (TMDs)
and their optical/optoelectronic properties \cite{Mounet_2018}. These
materials, which have been studied in their bulk form since the 1960's
\cite{1963,Fortin_1982}, have been shown to be good candidates for
various optical and optoelectronic applications \cite{lv_transition_2015,mueller_exciton_2018,wang_colloquium_2018,ma_tunable_2020}. 

The optical response of these materials is dominated by excitons \cite{nwu078},
and optically bright exciton absorption peaks have been shown to correspond
to the excitation of states in the $ns$ series \cite{schneider_two-dimensional_2018,hsu_dielectric_2019}.
As for $np$--series excitons, these can be controlled magnetically
despite being optically dark in TMDs \cite{zhang_magnetic_2017}. The selection rules of excitons in TMDs for absorption experiments have been recently thoroughly studied\cite{PhysRevB.99.235433}.

The linear dielectric response of TMDs consists of two distinct regimes:
the interband regime, where electrons from the valence band are excited
to the conduction band, leaving being a hole \cite{durnev_excitons_nodate,wang_colloquium_2018}.
The attractive electrostatic interaction between this newly--formed
electron--hole pair leads to the formation of a bound state (exciton),
and its $ns$ states are optically active and can be observed in absorption
measurements \cite{Potemski2017}. The second regime, characterized
by intra-exciton transitions, consists on the transition between the
excitonic ground state ($1s$) and the empty $np$ states \cite{berghauser_optical_2016,miyajima_optical_2016}.
Each of the $1s\rightarrow np$ transitions \cite{frohlich_observation_1985,berghauser_mapping_2018,tian_electronic_2020}
is characterized by a peak in the dynamical polarizability which,
in turn, determines the dielectric response of the system in a pump--probe
experiment \cite{poellmann_resonant_2015,Merkl2019}. 

\begin{figure}
\centering
\begin{centering}
\includegraphics[scale=1]{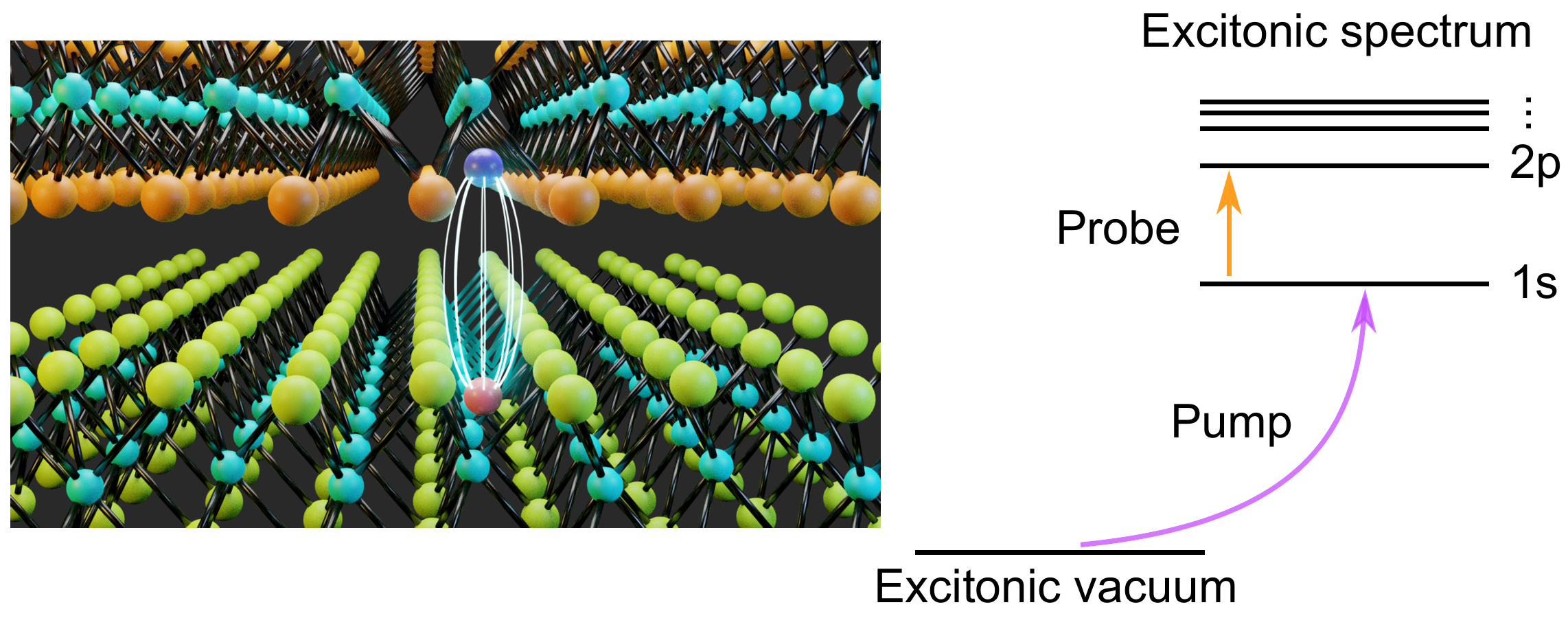}
\par\end{centering}
\caption{Left: Interlayer exciton in a van der Waals heterostructure. Right: Diagram of pump-probe experiments for observing $np$ (dark) excitonic states.}\label{fig:system}
\end{figure}

Interlayer excitons are formed when two monolayers are brought together
with type--II band alignment, where the conduction band minimum and
the valence band maximum are located in different layers \cite{Rivera2016,Jin2018}.
This allows the intralayer electron--hole pairs to tunnel into interlayer
excitons (Fig. \ref{fig:system}--left), which have a significantly longer life--time due to the
small overlap of the individual electron and hole wave functions \cite{Gong2014,Miller_2017}.
Additionally, interlayer excitons exhibit luminescence at lower energies
than their intralayer counterpart \cite{Vialla_2019} which, together
with their lower binding energies, allows for an easier identification
of the specific species in question in polarizability measurements
\cite{Merkl2019}. Recently, Merkl \textit{et al.}\cite{Merkl2019} were able to observe
the transition between interlayer and intralayer excitonic phases
in van der Waals heterostructures. This observation was performed via the measurement of the linear dielectric function of the excitons in a pump--probe experiment (Fig. \ref{fig:system}--right), observing a significant shift in maximum of the optical conductivity.

The excitonic energy levels are known to depend on whether the system
is confined or not (for example, the Rydberg series for the two--dimensional
Hydrogen atom is strongly affected by radial confinement \cite{chaoscador_two-dimensional_2005}).
Following recent advancements in the fabrication of TMD--based meta--materials
with atomic precision \cite{munkhbat_transition_2020,PhysRevB.103.L161401}, we focus our
attention on radially confined interlayer excitons in a $\mathrm{WSe}_{2}/\mathrm{WS}_{2}$
bilayer. 

From studying the dynamic polarizability of the excitonic states,
the full dielectric response of the excited
sample can be extracted. This transition from polarizability to dielectric response
is one from the microscopic (individual exciton) to the macroscopic
(when many excitons form a low--density exciton gas). The susceptibility of this exciton gas would be given, in a simplified form, as $\chi_{\mathrm{ex}}^{\mathrm{inter}}=\frac{N_{X}\alpha}{\epsilon_{0}}$,
with $N_{X}$ the exciton density in the two--dimensional material, $\alpha$ the linear polarizability, and $\epsilon_{0}$ the vacuum dielectric constant. Separating the real and imaginary part of the susceptibility, these can be experimentally measured \cite{wang_exciton_2006,poellmann_resonant_2015,cha_1_2016,Merkl2019}
and, therefore, would provide a good comparison point for the results we
obtain for the linear polarizability. 

The non--linear polarizabilities of the hydrogen atom have been thoroughly studied throughout the years \cite{Shelton1987,Manakov2004}, providing a good starting point for the study of these same non--linear processes in excitons. Recently, the non--linear optical effects from intra--excitonic transitions have been the focus of many theoretical and experimental works. This has been performed for several TMDs, including but not limited to $\mathrm{WSe}_{2}$ \cite{Zeng2013,Rosa2018} and $\mathrm{WS}_{2}$ \cite{Janisch2014,TorresTorres2016}, as well as $\mathrm{hBN}$ homostructures \cite{Yao2021}. Through doping processes, the optical non--linear properties of TMDs can be tuned \cite{PhysRevB.102.115310} which, along with their large non--linear optical coefficients \cite{PhysRevB.99.235433}, increases their feasibility for practical aplications \cite{Wang2015} (e.g., as optical modulators). 

We begin this paper by discussing the necessary modifications to the
Rytova--Keldysh potential \cite{rytova1967,keldysh1979coulomb} when
dealing with interlayer excitons, followed by analyzing an harmonic oscillator approximation to this potential which will motivate the chosen variational wave--function.
After defining this \textit{ansatz}, we outline the complete set of
basis functions we will use to approximate the excited states of the exciton.
The necessary material--dependent set of parameters characteristic
of the specific van der Waals heterostructure in question will also
be discussed, obtaining the desired energy levels. 

In Section \ref{sec:Variational-Polarizability} we outline Fowler's
and Karplus' method \cite{karplus_variationperturbation_1963,fowler_energy_1984}
to compute the dynamical polarizability of non--interacting
excitons. We then apply this same method to interlayer excitons in a $\mathrm{WSe}_{2}/\mathrm{WS}_{2}$--based
van der Waals heterostructure, comparing the obtained resonances against the transitions calculated both numerically and variationally via a finite--basis approach.

In Section \ref{sec:third-order}, we turn to the non--linear response of the excitons \cite{Mossman2016,Kocherzhenko2019}. We focus our discussion on the third--order polarizability, first outlining a variational procedure based on the ideas of both Karplus and Svendsen \cite{karplus_variationperturbation_1963,svendsen1988hyper,GABRYL1995199} for the calculation of the intra--excitonic third--order polarizability. This procedure was recently studied by Henriques \textit{et al.}, applied to for the study of the two--photon absorption for excitons in $\mathrm{WSe}_{2}$ \cite{henriques2021calculation}. After outlining this variational method, we look at the two--photon absorption for excitons in $\mathrm{WSe}_{2}/\mathrm{WS}_{2}$, discussing the various intra--excitonic transitions observed. Finally, we consider the third--harmonic generation, first with an extremely small broadening as to clearly differentiate each individual resonance, and afterwards with two much larger broadenings as to ascertain the feasibility of experimental detection of each peak. A diagrammatic representation of the various transitions is also presented for both the two--photon absorption and third--harmonic generation processes.

We finish the paper with our closing remarks in Section \ref{sec:Conclusions}.

\section{Interlayer Rytova--Keldysh Potential}

We begin this section by discussing the Hamiltonian in which
we focus our attention.This will lead to the discussion of the necessary modifications to the Rytova--Keldysh potential when
dealing with interlayer excitons. We define the basis for our variational
approach to the excitonic states (starting by the $np$--series states for Section \ref{sec:Variational-Polarizability} and then the $ns$-- and $nd$--series for Section \ref{sec:third-order}). We finish this section
by comparing the various variational energies for the excitonic ground state against the result from numerical integration of the Schr\"{o}dinger equation in a $\log$--grid \cite{johnson_new_1977,johnson_renormalized_1978,berghe_numerical_1989} via the Numerov shooting method.

\subsection{Model Hamiltonian in the Dipole Approximation}

Let us now begin by considering the following Hamiltonian (in atomic
units, as will be used throughout this paper)
in the dipole approximation
\begin{align}
H & =H_{0}-\mathbf{r}\cdot\mathbf{F}\left(t\right)\nonumber \\
 & =-\frac{1}{2\mu}\nabla^{2}+V\left(r\right)-\mathbf{r}\cdot\mathbf{F}\left(t\right),\label{eq:ham}
\end{align}
where $\mu$ is the reduced mass of the electron--hole system, $\nabla^{2}$
is the Laplacian operator (taken in polar coordinates throughout this paper), $V\left(r\right)$ is a potential energy
term and $\mathbf{F}\left(t\right)$ is an external time--dependent
field. This external field will be initially ignored as to first outline
the variational methods and wave functions that will be used, and
its action will be studied in Sections \ref{sec:Variational-Polarizability} and \ref{sec:third-order},
when the methods to obtain both the linear and the third--order dynamical polarizability are described, respectively. 

Considering this same electron--hole system in a van der Waals (vdW)
heterostructure, their potential energy is accurately modeled by the
Rytova--Keldysh potential \cite{rytova1967,keldysh1979coulomb}
\begin{equation}
V_{RK}\left(r\right)=-\frac{\pi}{2r_{0}}\left[\mathbf{H}_{0}\left(\kappa\frac{r}{r_{0}}\right)-\mathbf{Y}_{0}\left(\kappa\frac{r}{r_{0}}\right)\right],\label{eq:RK}
\end{equation}
with $\kappa$ the mean dielectric constant of the media, $r_{0}$
an intrinsic parameter of the 2D material (interpretable as an in--plane
screening length), and $\mathbf{H}_{0},\mathbf{Y}_{0}$ the Struve--H
and Bessel--Y (second kind) special functions of zero--th order,
respectively. The polarizability of intralayer excitons is discussed in detail in \cite{henriques2020optical,henriques2021calculation}.

When considering interlayer excitons, a minimum separation originating from the physical distance between the two layers appears
in the Rytova--Keldysh potential. The interlayer modified Rytova--Keldysh
potential then reads \cite{Kamban2020}
\begin{align}
V_{\mathrm{i-RK}}\left(r\right) & =-\frac{\pi}{2r_{0}}\left[\mathbf{H}_{0}\left(\kappa\frac{\sqrt{r^{2}+d^{2}}}{r_{0}}\right)-\right.\nonumber \\
& \left. \qquad- \mathbf{Y}_{0}\left(\kappa\frac{\sqrt{r^{2}+d^{2}}}{r_{0}}\right)\right],\label{eq:interlayer_RK}
\end{align}
with $d$ this interlayer separation distance. 

Presenting the same Coulomb tail behaviour as the Rytova--Keldysh
potential [Eq. (\ref{eq:RK})] at large distances, the finite interlayer separation $d$
eliminates the logarithmic divergence at the origin. The absence of
this divergence makes it so that an inverse--exponential \textit{ansatz}
for the excitonic wave function \cite{PhysRevB.94.125424} is not
the most adequate.  This, together with the parabolic nature of the interlayer modified Rytova--Keldysh potential near $r=0$, motivates the \textit{ansatz} that will be obtained in the
following section. 

\subsection{Gaussian--Based \textit{Ansatz}\label{sec_ansatz}}

Ahead we introduce a Gaussian wave function to describe the ground state of the interlayer $1s$ state analytically. In order to motivate this choice, we show below that the potential Eq. (\ref{eq:interlayer_RK}) can, indeed, be approximated by a parabolic potential near $r=0$.

In this regime, Eq.  (\ref{eq:interlayer_RK}) can be expanded up to second order in
$r$ as 
\begin{equation}
V\left(r\right)\approx-V_{0}+\gamma r^{2},
\end{equation}
where
\begin{align}
V_{0} & =\frac{\pi}{2r_{0}}\left[\mathbf{H}_{0}\left(\kappa\frac{d}{r_{0}}\right)-\mathbf{Y}_{0}\left(\kappa\frac{d}{r_{0}}\right)\right], \nonumber
\end{align}
\begin{align}
\gamma & =-\frac{\pi\kappa}{4dr_{0}^{2}}\left[\mathbf{H}_{1}\left(\kappa\frac{d}{r_{0}}\right)-\mathbf{Y}_{-1}\left(\kappa\frac{d}{r_{0}}\right)\right].
\end{align}
Considering this potential, the Hamiltonian will be
\[
H=-\frac{\hbar^{2}}{2\mu}\nabla^{2}-V_{0}+\gamma r^{2}.
\]

\begin{figure}
\centering
\begin{centering}
\includegraphics[scale=0.7]{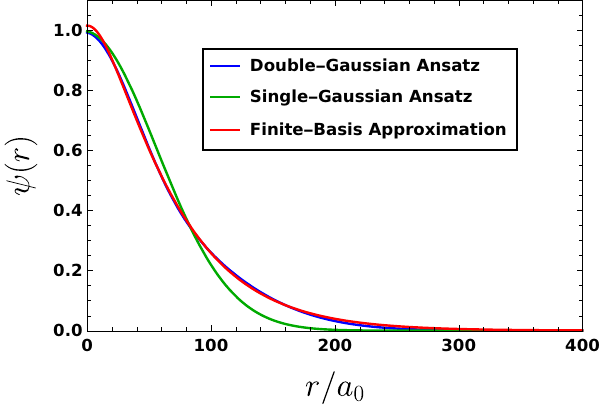}
\par\end{centering}
\caption{Radial profile of the Double--Gaussian Ansatz [Eq. (\ref{eq:gaussian_ansatz-1})],  the Gaussian Ansatz [Eq. (\ref{eq:gaussian_ansatz})] and the Finite--Basis Approximation (Sec. \ref{sec:besselJ-basis}) of the ground--state excitonic wave function.}\label{fig:wave_functions_ansatz}
\end{figure}

For the harmonic oscillator it is well known \cite{Edery_2018,cohen-tannoudji_quantum_2019,MartinsQuintela2020} that the $1s$ function is proportional to a Gaussian, that is, we have
\[
\Psi\left(\eta\right)\propto e^{-\eta/2},
\]
where $\eta=\mu\omega r^{2}/\hbar$ and $\omega=\sqrt{2\gamma/\mu}$. As such, an \textit{ansatz }based on this solution is given by
\begin{equation}
\psi\left(r\right)=\mathcal{C}e^{-\frac{r^{2}}{2\beta}},\label{eq:guass}
\end{equation}
where $\mathcal{C}$ is a normalization constant and $\beta$ is a
variational parameter. As we are interested in studying interlayer
excitons confined to a finite radius, a multiplicative factor is added
to impose boundary conditions, 
\begin{equation}
\psi_{0}\left(r\right)=\mathcal{C}e^{-\frac{r^{2}}{2\beta}}\left(R-r\right),\label{eq:gaussian_ansatz}
\end{equation}
where $R$ is the radius of the enclosure, considered $R=1200$ throughout this paper (in atomic units). This radius is considered sufficiently
large such that the variational wave functions become zero significantly
before the boundary is reached (clearly seen in Fig. \ref{fig:wave_functions_ansatz}, where the wave functions are already zero at around $\sim R/3$).

Similarly to the \textit{ansatz} defined by Pedersen in \cite{PhysRevB.94.125424},
we modify Eq. (\ref{eq:gaussian_ansatz}) into a sum of two gaussians, given by 
\begin{equation}
\psi_{0}\left(r\right)=\mathcal{C}\left(e^{-\frac{r^{2}}{2a}}+be^{-\frac{r^{2}}{2c}}\right)\left(R-r\right),\label{eq:gaussian_ansatz-1}
\end{equation}
which will the be \textit{ansatz} of the excitonic ground state we will consider throughout this article. Additionally, a visual comparison of Eq. (\ref{eq:gaussian_ansatz}) against Eq. (\ref{eq:gaussian_ansatz-1}), as well as the approximate wave function obtained via the method described in Sec. \ref{sec:besselJ-basis}, is present in Fig. \ref{fig:wave_functions_ansatz}. Clearly the \textit{ansatz} (\ref{eq:gaussian_ansatz-1}) performs much better than the \textit{ansatz} (\ref{eq:gaussian_ansatz}).

\subsection{Variational Approach from Boundary Conditions}\label{sec:besselJ-basis}

In certain conditions, namely for a small interlayer separation
$d$, the obtained eigenvalues from minimization of the Hamiltonian
\begin{equation}
H_{0}=-\frac{1}{2\mu}\nabla^{2}+V_{\mathrm{i-RK}}\left(r\right)\label{eq:unperturbed_ham}
\end{equation}
with the \textit{ansatz }from Eq. (\ref{eq:gaussian_ansatz-1}) are not significantly
different from those obtained from numerical integration (a small
difference is present for the present case in Table \ref{tab:Comparison-of-the-1}).
Additionally, obtaining higher energy states involves orthogonalization
against Eqs. (\ref{eq:gaussian_ansatz}) or (\ref{eq:gaussian_ansatz-1}),
followed by normalization, a process whose complexity increases substantially
when increasing the number of excited states \cite{grasselli_variational_2017}. 

A different approach to obtaining variational solutions to the Schr\"{o}dinger
equation with the interlayer modified Rytova--Keldysh is by integrating
the Hamiltonian matrix elements for a complete basis whose elements
obey the necessary boundary conditions. Truncating this basis and
diagonalizing the Hamiltonian matrix, we obtain a set of variational
wave functions that converge towards the real wave functions of the
system as the basis size increases.

Following \cite{Castano2005}, a basis of functions can be constructed
from the solutions of the circular infinite well, given by
\begin{equation}
\psi_{n,l}\left(r,\varphi\right)=\frac{C_{n,l}}{\sqrt{2\pi}}e^{il\varphi}J_{l}\left(z_{n,l}\frac{r}{R}\right),\label{eq:besselJ_basis}
\end{equation}
where $J_{l}\left(x\right)$ is the Bessel function of the first kind
of order $l$, $z_{n,l}$ is the $n$-th zero of $J_{l}\left(z\right)$, and $C_{n,l}$ is a normalization constant given by 
\[
C_{n,l}=\sqrt{\frac{2}{R^{2}J_{\left|l\right|+1}^{2}\left(z_{n,l}\right)}}.
\]

As both $V_{RK}\left(r\right)$ and $V_{i-RK}\left(r\right)$ are
invariant under rotations, the quantum number $l$ is well-defined.
As such, we can integrate the Hamiltonian and diagonalize it for a
finite number of functions with fixed $l$. Knowing this, we will
start the discussion by outlining the necessary material dependent
parameters for a $\mathrm{WSe}_{2}/\mathrm{WS}_{2}$ van der Waals
heterostructure.

\subsection{Material Dependent Parameters in $\mathrm{WSe}_{2}/\mathrm{WS}_{2}$}\label{subsec:-Parameters-1WSe2WS2}

To perform the necessary calculations and obtain the wave functions
and energy eigenvalues, we must substitute the material--specific
parameters characteristic of a $\mathrm{WSe}_{2}/\mathrm{WS}_{2}$
van der Waals (vdW) heterostructure \cite{Merkl2019}. 

The reduced mass of interlayer excitons in this vdW heterostructure
is $\mu=0.15$ and, as such, the effective Bohr radius will be $a_{0}\approx0.581\,\text{\AA}$.
The thickness of each layer is considered $d_{\mathrm{mat}}=5.7\,\text{\AA}$
and the gap distance is $d_{\mathrm{gap}}=1\,\text{\AA}$. As such,
the effective interlayer distance will be 
\begin{equation}
	d_{\mathrm{eff}}=\frac{1}{2}d_{\mathrm{mat}}+d_{\mathrm{gap}}+\frac{1}{2}d_{\mathrm{mat}}=6.7\,\text{\AA}.\label{eq:interlayer_distance}
\end{equation}

The permittivities for each material are given by 
\begin{equation}
\epsilon_{\left(n\right)} =\sqrt{\epsilon_{\left(n\right)}^{\bot}\epsilon_{\left(n\right)}^{\Vert}},
\end{equation} 
where $\epsilon_{\left(n\right)}^{\Vert}$ and $\epsilon_{\left(n\right)}^{\bot}$ are the in- and out-of-plane relative permittivities of the material $n$. The numerical values of each component of the relativie permittivities in each material is present in Table \ref{tab:dielectric_comparison}.
\begin{table}
\centering
\begin{centering}
\begin{tabular}{|c|c|c|}
\hline 
 & \textbf{$\mathrm{WSe_{2}}$} & \textbf{$\mathrm{WS_{2}}$} \tabularnewline
\hline\hline 
\textbf{$\epsilon_{\left(n\right)}^{\bot}$} & $7.5$ & $6.3$ 
\tabularnewline
\hline
\textbf{$\epsilon_{\left(n\right)}^{\Vert}$} & $13.36$ & $11.75$ 
\tabularnewline
\hline 
\hline
\textbf{$\epsilon_{\left(n\right)}$} & $10.01$ & $8.604$ 
\tabularnewline
\hline 
\end{tabular}
\par \end{centering}
\caption{\label{tab:dielectric_comparison}In--plane, out--of--plane, and total relative permittivities of $\mathrm{WSe}_{2}$ and $\mathrm{WS}_{2}$.}
\end{table}
Knowing the permittivities in each material, the average permittivity of the vdW heterostructure is 
\[
\kappa_{\mathrm{mat}}=\frac{\epsilon_{\mathrm{WSe_{2}}}+\epsilon_{\mathrm{WS_{2}}}}{2}=9.31.
\]

Regarding the screening length $r_{0}$, this parameter can be separated
as a sum for each layer (as described in \cite{Kamban2020}) 
\[
r_{0}\rightarrow r_{0}^{\left(1\right)}+r_{0}^{\left(2\right)},
\]
where $r_{0}^{\left(n\right)}$ denotes the screening length for the
layer $n$. The screening length for each individual layer can be obtained as \cite{poellmann_resonant_2015}
\begin{equation}
r_{0}^{\left(n\right)}=\frac{\epsilon_{n}^{2}-1}{2\epsilon_{n}}d_{n}\sqrt{\frac{\epsilon_{n}^{\Vert}}{\epsilon_{n}^{\bot}}},
\end{equation}
where $d$ is the thickness of the layer. As the in- and out-of-plane
relative permittivities of the two materials are known, the total
screening length is 
\begin{align*}
r_{0} & =70.73\,\text{\AA}.
\end{align*}

Choosing a basis size of $120$ and the same enclosure radius
as in Sec. \ref{sec_ansatz} ($R=1200$), the ground--state energy eigenvalue is (in both atomic units and meV)
\begin{equation*}
E_{1,0} =-0.00143378 = -39.0152\,\mathrm{meV}.
\end{equation*}
A comparison against both variational \textit{ansatze} [Eqs. (\ref{eq:gaussian_ansatz}) and (\ref{eq:gaussian_ansatz-1})] and the 
numerical results from considering the shooting method in a log--grid
is given in Table \ref{tab:Comparison-of-the-1}. 

\begin{table}
\centering
\begin{centering}
\begin{tabular}{|c|c|}
\hline 
 & \textbf{$1s$ State} \tabularnewline
\hline\hline 
\textbf{Finite Basis} & $-0.00143378$\tabularnewline
\hline
\textbf{Shooting Method} & $-0.00144375$\tabularnewline
\hline
\textbf{Double--Gaussian [Eq. (\ref{eq:gaussian_ansatz-1})]} & $-0.00142879$ \tabularnewline
\hline
\textbf{Gaussian [Eq. (\ref{eq:gaussian_ansatz})]} & $-0.00136314$\tabularnewline
\hline 
\end{tabular}
\par \end{centering}
\caption{\label{tab:Comparison-of-the-1}Comparison of the different estimates for the excitonic ground--state energy eigenvalue for a $\mathrm{WSe}_{2}/\mathrm{WS}_{2}$ heterostructure. The interlayer distance is given in Eq. (\ref{eq:interlayer_distance})}
\end{table}

\section{Dynamical Variational Method for Linear Polarizability}\label{sec:Variational-Polarizability}

In this section we will briefly outline Fowler's and Karplus' \cite{karplus_variationperturbation_1963,fowler_energy_1984}
method to compute the dynamical polarizability of various systems.
This discussion will be along the same lines as in \cite{henriques2020optical}, serving as a quick outline of the procedure as some results will be necessary further ahead.
We apply this method to calculate the linear polarizability of interlayer excitons in a $\mathrm{WSe}_{2}/\mathrm{WS}_{2}$ circular dot, comparing the obtained peaks with the
calculated $1s\rightarrow np$ transitions.

Looking back to Eq. (\ref{eq:ham}), we consider (without loss of generality)
the external field pointing along the $-x$ direction. Closely following the approach presented in \cite{henriques2020optical},
first outlined by Fowler and Karplus \cite{karplus_variationperturbation_1963,fowler_energy_1984}, the time--dependent Schr\"{o}dinger equation
reads 
\begin{equation}
\left[H_{0}+xF\left(t\right)\right]\left|\psi\left(t\right)\right\rangle =i\frac{\partial}{\partial t}\left|\psi\left(t\right)\right\rangle .\label{eq:time-dependent}
\end{equation}
where $\left|\psi\left(t\right)\right\rangle $ describes the wave
function of the system in the presence of the electric field $\mathbf{F}$.
The dynamical polarizability is defined as 
\begin{equation}
\alpha\left(\omega\right)=-\left\langle \psi_{0}\right|x\left|\psi_{1}^{+}\right\rangle -\left\langle \psi_{0}\right|x\left|\psi_{1}^{-}\right\rangle \label{eq:Fowler alpha}
\end{equation}
with $\psi_{0}$ the Gaussian \textit{ansatz} of Eq. (\ref{eq:gaussian_ansatz-1}) and $ \psi_{1}^{\pm} $ the excited $ p $--series states, such that
\begin{equation}
E\left(\omega\right)=E_{0}-\frac{1}{2}\alpha\left(\omega\right)F^{2}.
\end{equation}

As we are working on a finite disk of radius $R$, Bessel functions
of the first kind are an appropriate complete set of functions to
describe a problem in such a geometry, as defined in Eq. (\ref{eq:besselJ_basis}). Using these functions as a
basis, we write $\psi_{1}^{\pm}\left(\mathbf{r}\right)$ as
\begin{equation}
\psi_{1}^{\pm}\left(\mathbf{r}\right)=\cos\theta\sum_{n=1}^{N}c_{n}^{\pm}J_{1}\left(\frac{z_{1,n}r}{R}\right),\label{eq:psi_1 J_1}
\end{equation}
where $J_{1}\left(z\right)$ is the Bessel function of the first kind
of order $1$, $z_{1,n}$ is the $n$--th zero of $J_{1}\left(z\right)$,
$N$ is the number of Bessel functions we choose to use, and $\left\{ c_{n}^{\pm}\right\} $
are a set of coefficients yet to be determined. As proposed in Refs.
\cite{karplus_variationperturbation_1963,montgomery_one-electron_1978,yaris_timedependent_1963,yaris_timedependent_1964},
the values of $\left\{ c_{n}^{\pm}\right\} $ are determined from
the minimization of the functional
\begin{align}
\mathcal{J}_{\pm} & =\int d\mathbf{r}\,\psi_{1,n}^{\pm}\left(\mathbf{r}\right)\left[H_{0}-E_{0}\pm\hbar\omega\right]\psi_{1,n}^{\pm}\left(\mathbf{r}\right)+ \nonumber \\
&+2\int d\mathbf{r}\,\psi_{1,n}^{\pm}\left(\mathbf{r}\right)r\cos\theta\psi_{0}\left(\mathbf{r}\right),
\end{align}
The discussion of the minimization of the $\mathcal{J}$ functional and the computation of the $c_{n}^{\pm}$ coefficients is performed in Appendix \ref{app:A}.

The dynamical polarizability is then computed by substituting Eq. (\ref{eq:psi_1 J_1}) into Eq. (\ref{eq:Fowler alpha}), together with the $c_{n}^{\pm}$ coefficients discussed in Appendix \ref{app:A}, being written as
\begin{align}
\alpha\left(\omega\right) & =-g_{v}\pi\sum_{n=1}^{N}\left(c_{n}^{+}+c_{n}^{-}\right) \times \nonumber \\
& \qquad \times \int_{0}^{R}J_{1}\left(\frac{z_{1n}r}{R}\right)r\,\psi_{0}\left(r\right)r\,dr.\label{eq:polarizability-1}
\end{align}
where $g_{v}=2$ is the valley degeneracy. The $\omega$ dependence
on the right hand side is present in the coefficients $c_{n}^{\pm}$ [more explicitly, this dependence is present in the $\mathbb{M}$ matrix (Eq. \ref{eq:c_coeffs}) from the $g_{i}^{\pm}\left(\omega\right)$  term, as defined in Eq. (\ref{eq:g_vec})]. We remark that choosing for the ground state wave function a variational function allows us to write Eq. (\ref{eq:polarizability-1}) exclusively in terms of analytical functions. This considerably simplifies the numerical minimization.

\begin{figure}
\centering
\begin{centering}
\includegraphics[scale=0.7]{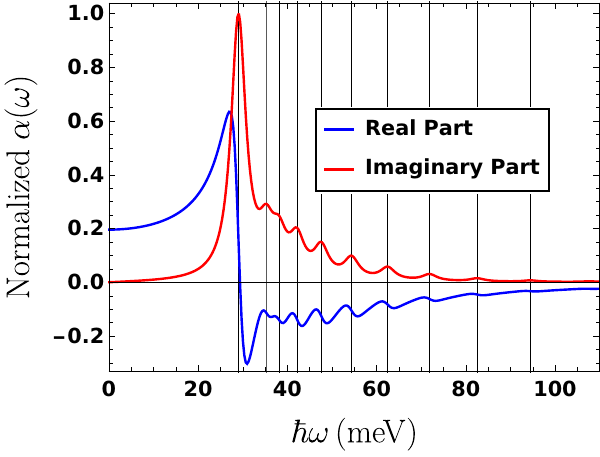}
\par\end{centering}
\caption{Real (blue) and imaginary (red) parts of the normalized polarizability for interlayer excitons in a $\mathrm{WSe}_{2}/\mathrm{WS}_{2}$ heterostructure.
The vertical black lines represent the excitonic $1s\rightarrow np$ transitions.}\label{fig:real_dielectric}
\end{figure}

The polarizability for interlayer excitons in a $\mathrm{WSe}_{2}/\mathrm{WS}_{2}$ heterostructure is plotted in Fig. \ref{fig:real_dielectric}, normalized by its value at the maximum for the excitonic $1s\rightarrow 2p$ transition (with $E_{1s\rightarrow 2p}\sim 29.0576\,\mathrm{meV}$) for easier visualization. Additionally, a brief comparison of the linear polarizability for interlayer and intralayer excitons is made in Appendix \ref{app:B}.

Having outlined and obtained the linear polarizability in Fig. \ref{fig:real_dielectric}, we now turn our attention to the third--order polarizability of interlayer excitons in this same vdW heterostructure. 

\section{Svendsen's Method for the Third--Order Polarizability}\label{sec:third-order}

Having outlined Fowler's and Karplus' method in Section \ref{sec:Variational-Polarizability}, we will now consider a similar method for the third--order polarizability, following the procedure delineated  by Svendsen \cite{svendsen1988hyper,GABRYL1995199}
and recently applied by Henriques \textit{et al} \cite{henriques2021calculation}
to intralayer excitons in TMDs.

Maintaining the definition for the $\mathbf{c}^{\pm}$ coefficients
given in Eq. (\ref{eq:c_coeffs}), and following closely \cite{GABRYL1995199,karplus_variationperturbation_1963,henriques2021calculation},
we write the functional 
\begin{align}
& \mathcal{K} =\left\langle \xi_{\alpha\beta}\left(\omega_{a},\omega_{b}\right)\middle|H_{0}-E_{0}+\omega_{a}+\omega_{b}\middle|\xi_{\alpha\beta}\left(\omega_{a},\omega_{b}\right)\right\rangle +\nonumber \\
 & \;+\left\langle \xi_{\alpha\beta}\left(\omega_{a},\omega_{b}\right)\middle|\mathbf{d}_{\beta}\middle|\psi_{\alpha}\left(\omega_{a}\right)\right\rangle+\left\langle \psi_{\alpha}\left(\omega_{a}\right)\middle|\mathbf{d}_{\beta}\middle|\xi_{\alpha\beta}\left(\omega_{a},\omega_{b}\right)\right\rangle \label{eq:functionalK}
\end{align}
where $\mathbf{d}$ is the dipole operator defined previously, $E_{0}$
the ground--state energy, $H_{0}$ the unperturbed Hamiltonian [defined
in Eq. (\ref{eq:unperturbed_ham})] and $\left|\psi_{\alpha}\right\rangle $
the basis obtained via the $\mathbf{c}^{\pm}$ coefficients and the
Bessel functions $J_{1}\left(z_{1,n}\right)$ in Eq. (\ref{eq:psi_1 J_1}).
This functional is then minimized with respect to $\left|\xi_{\alpha\beta}\right\rangle $,
defined analogously to Eq. (\ref{eq:psi_1 J_1}) as
\begin{align}
&\xi_{\alpha\beta}\left(\omega_{a},\omega_{b};r\right)= \\
&=\sum_{n=1}^{+\infty}\sum_{\left|l\right|=0}^{+\infty}\zeta_{l,n}^{\alpha,\beta}\left(\omega_{a},\omega_{b}\right)J_{l}\left(z_{l,n}\frac{r}{R}\right)\frac{e^{il\theta}}{\sqrt{2\pi}},\nonumber
\end{align}
where we will focus our attention on the $xx$ component of $\xi_{\alpha\beta}$. 

The discussion of minimization of this $\mathcal{K}$ functional, done in a similar fashion as for the $\mathcal{J}$ functional, is performed in-depth in Appendix \ref{app:C}.

Having defined both $\mathbf{\zeta}_{0}^{x,x}$ and $\mathbf{\zeta}_{2}^{x,x}$ in Eqs. (\ref{eq:zeta_l0}--\ref{eq:zeta_l2}), respectively, the third--order susceptibility follows from \cite{henriques2021calculation} (with $\omega_{\sigma}=\omega_{1}+\omega_{2}+\omega_{3}$) as
\begin{widetext}
\begin{align}
\chi_{x,x,x,x}^{\left(3\right)}\left(-\omega_{\sigma};\omega_{1},\omega_{2},\omega_{3}\right) & =\frac{1}{3 !} \mathcal{P}\left\lbrace-\mathbf{c}^{\dagger}\left(\left(-\omega_{\sigma}\right)^{\dagger}\right)\cdot\left[\mathbb{T}_{0} \cdot \zeta_{0}^{x,x}\left(-\omega_{2},-\omega_{3}\right)+\mathbb{T}_{2} \cdot \zeta_{2}^{x,x}\left(-\omega_{2},-\omega_{3}\right)\right]\right.+ \nonumber \\
& \left.+\left[ \mathbf{S} \cdot \mathbf{c}\left(-\omega_{\sigma}\right) \right] \times\left[\sum_{n=1}^{N}c_{n}^{\dagger}\left(\left(-\omega_{2}\right)^{\dagger}\right)c_{n}\left(\omega_{1}\right)2\pi\frac{R^{2}}{2}\left[J_{2}\left(\mathrm{z}_{1, n}\right)\right]^{2}\right]\right\rbrace,\label{eq:tomod}
\end{align}
\end{widetext}
with $\mathcal{P}$ denoting the different permutations of the  frequencies $\left(-\omega_{\sigma};\omega_{1},\omega_{2},\omega_{3}\right)$, $\mathbb{T}_{0/2}$ is defined in Eq. (\ref{eq:T_matrix}), and $\mathbf{S}$ is the vector in Eq. (\ref{eq:S integral}).
Having defined Eq. (\ref{eq:tomod}), we will now utilize it to compute two different regimes for the third--order polarizability of interlayer excitons in $\mathrm{WSe}_{2}/\mathrm{WS}_{2}$.

\subsection{Two--Photon Absorption in $\mathrm{WSe}_{2}/\mathrm{WS}_{2}$}\label{sec:TPA}
\begin{figure*}
\centering
\begin{centering}
\includegraphics[scale=0.7]{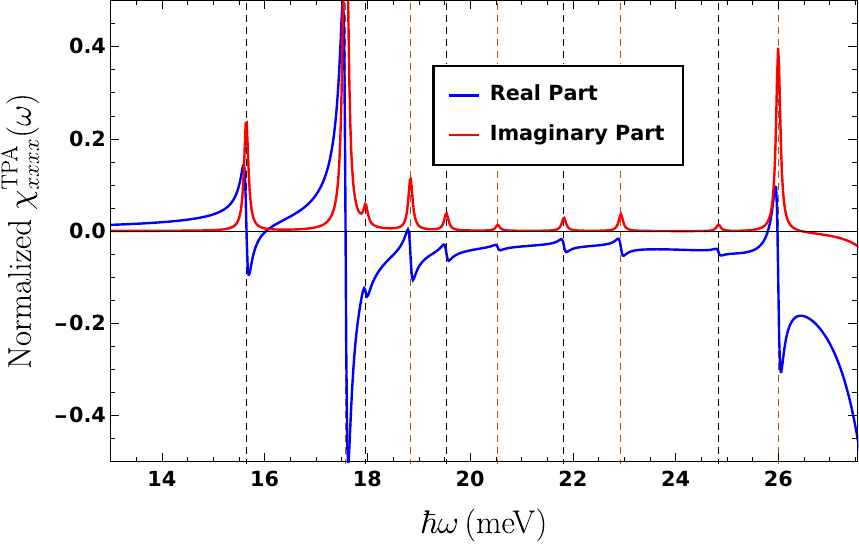}
\includegraphics[scale=0.6]{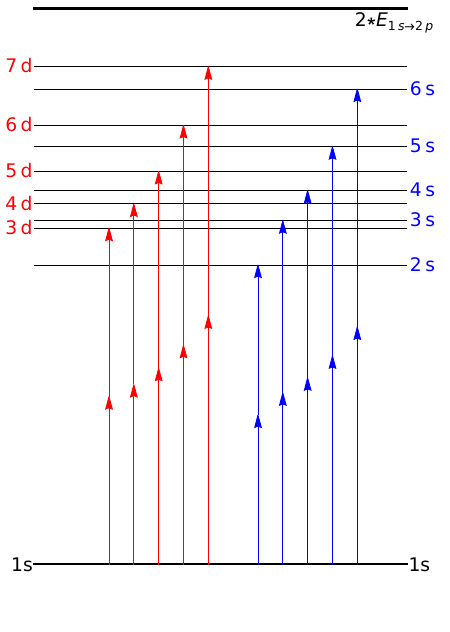}
\end{centering}
\caption{\\Left: Real (blue) and imaginary (red) parts of the two--photon absorption third--order susceptibility (normalized by the peak value for the $1\rightarrow 3d$ transition at $\sim17.5\,\mathrm{meV}$) as a function of the incident photons' energies for a broadening $\Delta=0.05\,\mathrm{meV}$ and basis size $N=120$. The resonances correspond to $1s\rightarrow ns$ transitions (black dashed lines) and to $1s\rightarrow nd$ transitions (dark–orange dashed lines). The dominant $1s\rightarrow2p$ transition is at $\sim29.0576\,\mathrm{meV}$, outside of the frequency range plotted in this figure. Right: Diagramatic representation of the $1s\rightarrow ns$ (blue arrows) and $1s\rightarrow nd$ (red arrows) transitions. Topmost black line represents the resonance when $\hbar\omega=E_{1s\rightarrow2p}$. We remark that the red and blue arrows represent the selection rules of the system, that is, the only allowed transitions.}\label{fig:TPA}
\end{figure*}
We will now compute the $xxxx$ component of the two--photon absorption third--order susceptibility, defined as $\chi_{xxxx}^{\mathrm{TPA}}\left(\omega\right)=\chi_{xxxx}^{\left(3\right)}\left(-\omega;\omega,-\omega,\omega\right)$.

As the two--photon absorption was considered in \cite{henriques2021calculation}, we will not discuss it as in--depth as we will with the third--harmonic generation. Similarly to what was done in Fig. \ref{fig:real_dielectric}, we plot the real and imaginary parts of the two--photon absorption third--order susceptibility for a broadening small enough $\left(0.05\,\mathrm{meV}\right)$ such that each individual peak can be clearly observed. This plot is visible in Fig. \ref{fig:TPA}, normalized by the peak value for the $1\rightarrow 3d$ transition at $\sim16.6\,\mathrm{meV}$, along with a diagram representing the $1s\rightarrow ns$ and $1s\rightarrow nd$ transitions.

\subsection{Third--Harmonic Generation}\label{sec:THG}

To finalize this paper, we will now compute the $xxxx$ component of the third--order susceptibility third--harmonic generation of interlayer excitons in $\mathrm{WSe}_{2}/\mathrm{WS}_{2}$, which is defined as $\chi_{xxxx}^{\mathrm{THG}}\left(3\omega\right)=\chi_{xxxx}^{\left(3\right)}\left(-3\omega;\omega,\omega,\omega\right)$. Following the same approach as in Sec. \ref{sec:TPA}, we will begin by computing $\chi_{xxxx}^{\mathrm{THG}}\left(3\omega\right)$ for a small enough broadening such that each resonance is clear. 

\begin{figure*}
\centering
\begin{centering}
\includegraphics[scale=0.7]{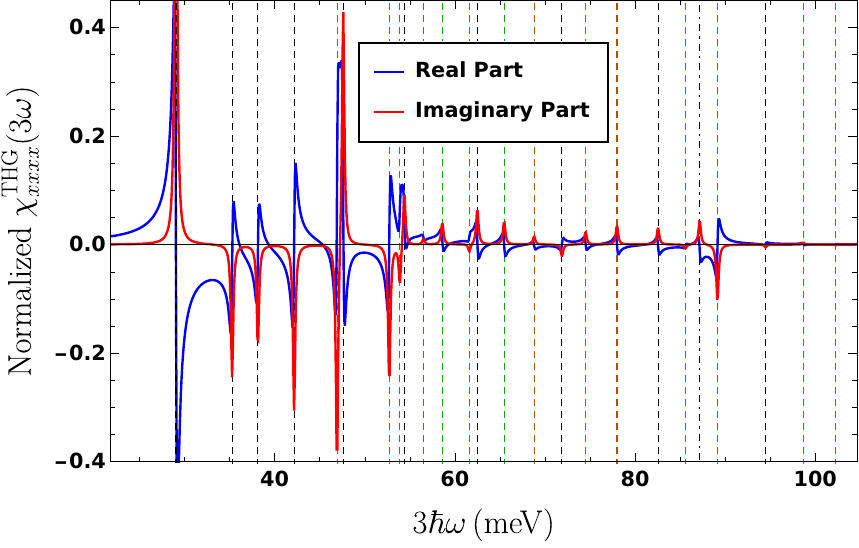}\includegraphics[scale=0.6]{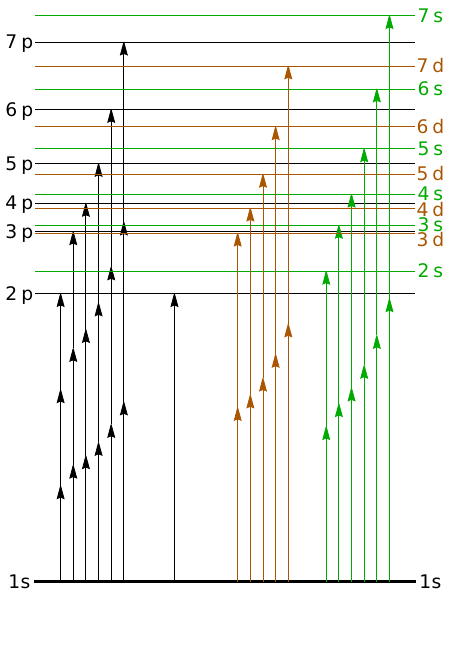}
\end{centering}
\caption{\\Left: Real (blue) and imaginary (red) parts of the third--harmonic generation third--order susceptibility (normalized by the peak value for the $1\rightarrow 2p$ resonance at $\sim 29.0576,\mathrm{meV}$) as a function of the incident photons' energies for a broadening $\Delta=0.05\,\mathrm{meV}$ and basis size $N=120$. The resonances correspond to $1s\rightarrow np$ transitions (black dashed lines), $1s\rightarrow ns$ transitions (green dashed lines) and $1s\rightarrow nd$ transitions (dark--orange dashed lines). The $\sim 87.1727\,\mathrm{meV}$ resonance (black dot--dashed line) corresponds to the response at frequency $3\hbar\omega=3E_{1s\rightarrow2p}$. Right: Diagramatic representation of the various transitions visible in the plot on the left--hand side. The colors of the arrows represent the different transitions: $1s\rightarrow np$ (black), $1s\rightarrow ns$ (green), and $1s\rightarrow nd$ (dark--orange), while the number of arrows represents the frequency of the transition: 3 arrows for $3\hbar\omega = E_{i\rightarrow f}$, 2 arrows for $2\hbar\omega = E_{i\rightarrow f}$, and 1 arrow for $\hbar\omega = E_{i\rightarrow f}$. Again, as in Fig. \ref{fig:TPA}, the arrows represent the selection rules associated with the third--harmonic generation (THG) nonlinear process. }\label{fig:THG}
\end{figure*}

The left--hand side of Fig. \ref{fig:THG} has been normalized by the peak value for the resonance at $3\hbar\omega=E_{1s\rightarrow 2p}$, whilst the right--hand side features a diagram representing the various transitions. 

\begin{figure*}
\centering
\begin{centering}
\includegraphics[scale=0.7]{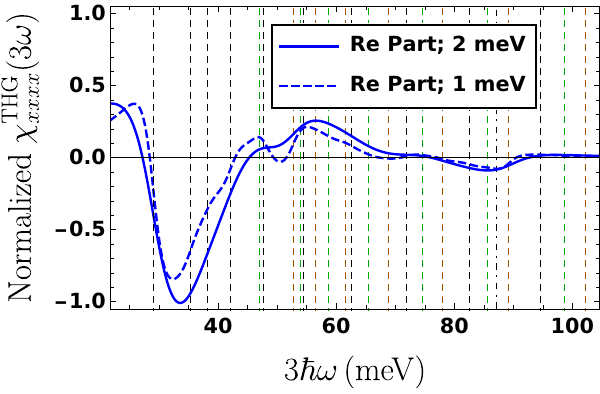}\;\includegraphics[scale=0.7]{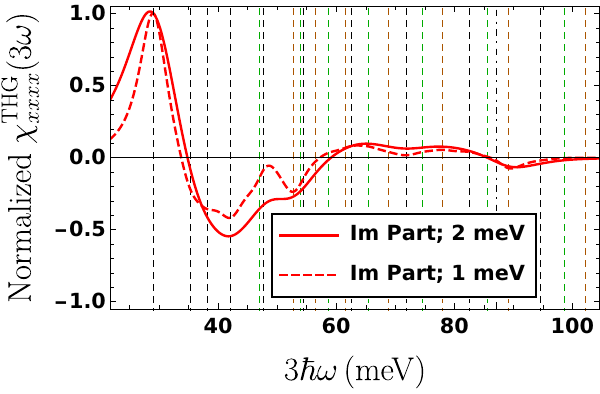}
\end{centering}
\caption{Real (left) and imaginary (right) parts of the normalized third--harmonic generation third--order susceptibility for different broadening values: $\Delta = 2\,\mathrm{meV}$ (solid lines) and $1\,\mathrm{meV}$ (dashed lines). The basis size is fixed at $N=120$. The resonances correspond to $1s\rightarrow np$ transitions (black dashed lines), $1s\rightarrow ns$ transitions (green dashed lines) and $1s\rightarrow nd$ transitions (dark–orange dashed lines). The $\sim 87.1727\,\mathrm{meV}$ resonance (black dot--dashed line) corresponds to the response at  $3\hbar\omega=3E_{1s\rightarrow2p}$.}\label{fig:THG_comp}
\end{figure*}

Spectral broadening as low as $\Delta\sim 2\,\mathrm{meV}$ (i.e., the broadening that was considered for Fig. \ref{fig:real_dielectric}) can be achieved for low temperature encapsulated systems \cite{Robert2018}. Considering much higher values of the broadening than in Fig. \ref{fig:THG}, namely $\Delta=1\,\mathrm{meV}$ and $\Delta=2\,\mathrm{meV}$, we recompute $\chi_{xxxx}^{\mathrm{THG}}\left(3\omega\right)$, obtaining Fig. \ref{fig:THG_comp}.

Although some $1s\rightarrow np$ resonances are already apparent at $\Delta=2\,\mathrm{meV}$, namely at $\sim 40\,\mathrm{meV}$ ($1s\rightarrow 5p$) and $\sim 60\,\mathrm{meV}$ ($1s\rightarrow 8p$), most $1s\rightarrow ns$ and  $1s\rightarrow nd$ transitions only become clearly visible in the sub--meV regime. At those broadenings, the most clear peaks are at $3\hbar\omega\approx 50\,\mathrm{meV}$ ($1s\rightarrow 3d$) and at $3\hbar\omega\approx 56\,\mathrm{meV}$ ($1s\rightarrow 4s$). The two resonances at $\sim 72\,\mathrm{mev}$ and $\sim 75\,\mathrm{mev}$ are joined into one at $\Delta = 1\,\mathrm{meV}$, although their presence is noticeable from comparison with the adjacent $1s\rightarrow np$ transition peaks.

The $1s\rightarrow np$ resonance peaks occur for $3\hbar\omega=E_{1s\rightarrow np}$, while the $1s\rightarrow ns$ and  $1s\rightarrow nd$ peaks are situated at $3\hbar\omega=\frac{3}{2}E_{1s\rightarrow ns/nd}$. Additionally, the peak at $3\hbar\omega= 87.1727\,\mathrm{meV}$ corresponds to $3\hbar\omega=3E_{1s\rightarrow2p}$. This is due to the response at both $2\omega$ \cite{acs.nanolett.0c03763} and $\omega$  also being present in this process, described in detail in Section 3.2 of \cite{boyd2020nonlinear}, i.e. $2\hbar\omega=E_{1s\rightarrow ns/nd}$ and $\hbar\omega=E_{1s\rightarrow2p}$. No other peaks for $\hbar\omega=E_{1s\rightarrow np}$ are visible, as they would already be outside the considered frequency domain ($3E_{1s\rightarrow3p}\sim 105.872\,\mathrm{meV}$, barely outside of the plot).

\section{Conclusions}\label{sec:Conclusions}

In this paper, we discussed interlayer excitons in a $\mathrm{WSe}_{2}/\mathrm{WS}_{2}$--based van der Waals
heterostructure, for which recent experimental measurements of the polarizability through pump--probe
experiments \cite{Merkl2019} have been performed. In these experiments,
the transition from bright to dark states has been accessed for both
interlayer and intralayer excitons. In this paper, we focused ourselves
solely on the transitions between intra--excitonic states for interlayer excitons in a circular $\mathrm{WSe}_{2}/\mathrm{WS}_{2}$ dot. 

Approximating the interlayer modified Rytova--Keldysh potential in a power series, we motivate a Gaussian variational \textit{ansatz} for the ground state of the system. 
After discussing the necessary material--dependent parameters, we consider both this \textit{ansatz} and a complete set of basis functions the Bessel functions of the first kind to estimate the excitonic wave functions through variational methods. 

We apply Fowler's and Karplus' method variational method to access the linear polarizability of two--dimensional interlayer excitons, focusing on the transition from bright to dark states (more specifically, $1s\rightarrow np$ transitions) in a circular dot of a $\mathrm{WSe}_{2}/\mathrm{WS}_{2}$--based van der Waals heterostructure with Dirichlet boundary conditions. 

For the linear regime we consider a broadening of $\Delta=2\,\mathrm{meV}$, allowing for the clear observation of the multiple peaks in the excitonic polarizability. We observe an almost perfect agreement between the frequency of each resonance and the energy differences between the ground state and the excited states of the exciton. These energy differences were calculated both numerically (via the Numerov shooting method) and variationally (via both a Gaussian \textit{ansatz} for the ground--state and a finite basis of Bessel function of the first kind), allowing for a greater confidence in their values. As expected \cite{Merkl2019}, comparing the linear polarizability of interlayer excitons in the hetero--bilayer against intralayer excitons in each of the individual layers shows that the binding energies are much lower in the interlayer regime (Figs. \ref{fig:real_dielectric} and \ref{fig:inter-intra_comparison}, Table \ref{tab:Comparison-of-the-1} and Eq. (\ref{eq:monolayer_energies})).

Onto the third--order polarizability, we began by providing an in--depth discussion of Svendsen's variational method \cite{svendsen1988hyper,GABRYL1995199}. After outlining this method and analyzing the required modifications due to the necessity of orthogonality between the various states, we work on the algebraization of the method, arriving at a purely vectorial problem after the necessary integrals are computed. The obtained expression is then tested first with the two--photon absorption, and then with third--harmonic generation.

Starting by two--photon absorption, a process which was discussed in--depth by Henriques \textit{et al.} \cite{henriques2021calculation}, we perform the necessary calculations with a broadening small enough such that the resonance associated with each individual intra--excitonic energy level transition can be clearly identifiable $\left( \Delta=0.05\,\mathrm{meV}\right)$. The two--photon absorption third--order susceptibility is then normalized by the value at the $E_{1s\rightarrow 3d}$ resonance as to facilitate the comparison of each peak. Of note is the amplitude of the resonance associated with the $E_{1s\rightarrow 7d}$ transition, peaking at around $40\%$ of the maximum and being dramatically larger than the adjacent transitions. Additionally, the individual selection rules for this system were identified.

Finally turning to the third--harmonic generation in the third--order susceptibility, we first perform the calculations for a very small broadening $\left( \Delta=0.05\,\mathrm{meV}\right)$. Each individual transition was clearly identifiable, with resonances at $\hbar \omega = E_{i\rightarrow f}$ and $3\hbar \omega = E_{i\rightarrow f}$ for $1s \rightarrow np$ transitions, and at $2\hbar \omega = E_{i\rightarrow f}$ for both $1s \rightarrow ns$ and $1s \rightarrow nd$ transitions. The individual selection rules for the third--harmonic generation process were clearly identifiable.

Considering experimentally--obtainable values of the broadening, we compare the third--harmonic generation third--order susceptibility for  $\Delta = 1\,\mathrm{meV}$ and $2\,\mathrm{meV}$. We argue for the feasibility of experimentally observing the three well--resolved resonance peaks, associated with $1s\rightarrow 2p$, $1s\rightarrow 3d$, and $1s\rightarrow 4s$ transitions, where the latter two could be an experimental indication of third--harmonic generation due to intra--excitonic transitions in this specific van der Waals heterostructure.

\section*{Acknowledgments}

M.F.C.M.Q. acknowledges the International Nanotechnology Laboratory for the Quantum Portugal Initiative grant SFRH/BD/151114/2021. N.M.R.P acknowledges support by the Portuguese Foundation for Science and Technology (FCT) in the framework of the Strategic Funding UIDB/04650/2020. J.C.G.H. acknowledges the Center of Physics for a grant funded by the UIDB/04650/2020 strategic project. N.M.R.P. also acknowledges support from the European Commission through the project “Graphene--Driven Revolutions in ICT and Beyond” (Ref. No. 881603, CORE 3), COMPETE 2020, PORTUGAL 2020, FEDER and the FCT through projects POCI-01-0145-FEDER- 028114, POCI-01-0145-FEDER-02888 and PTDC/NAN-OPT/29265/2017.

\appendix
\section{Variational Coefficients for the Linear Polarizability \label{app:A}}
Recalling the orthogonality relation of Bessel functions on a
finite disk of radius $R$ \cite{zwillinger_table_2014},
\begin{align}
\int_{0}^{R}J_{\nu}\left(\frac{z_{vm}r}{R}\right)J_{\nu}\left(\frac{z_{vn}r}{R}\right)r dr = & \nonumber \\
=\frac{R^{2}}{2}\delta_{nm}\left[J_{\left|\nu\right|+1}\left(z_{vm}\right)\right]^{2},&
\end{align}
with $z_{vm}$ the $m-$th zero of $J_{\nu}\left(z\right)$, one easily shows that the functional
can be rewritten as
\begin{align}
&\mathcal{J}_{\pm} =\pi\sum_{n=1}\sum_{k=1}c_{n}^{\pm}c_{k}^{\pm}\mathcal{I}_{kn}+2\pi\sum_{n=1}c_{n}^{\pm}\mathcal{S}_{n}+\nonumber \\
& +\frac{\pi R^{2}}{2}\sum_{n=1}\left(c_{n}^{\pm}\right)^{2}\left[\frac{\hbar^{2}}{2\mu}\frac{z_{1n}^{2}}{R^{2}}-E_{0}\pm\hbar\omega\right]\left[J_{2}\left(z_{1n}\right)\right]^{2},
\end{align}
where $\mathcal{I}_{kn}$ and $\mathcal{S}_{n}$ refer to the following
integrals involving two and one Bessel functions, respectively
\begin{equation}
\mathcal{I}_{kn} =\int_{0}^{R} J_{1}\left(\frac{z_{1k}r}{R}\right)V_{{\rm RK}}\left(r\right)J_{1}\left(\frac{z_{1n}r}{R}\right)r\, dr,\label{eq: I integral}
\end{equation}
\begin{equation}
\mathcal{S}_{n} =\int_{0}^{R} J_{1}\left(\frac{z_{1n}r}{R}\right)\,r\,\psi_{0}^{{\rm fin.}}\left(r\right)r dr.\label{eq:S integral}
\end{equation}

Differentiating $\mathcal{J}_{\pm}$
with respect to the different $c_{n}^{\pm}$
coefficients, one finds
\begin{align}
&c_{j}^{\pm}\left\{ \frac{R^{2}\left[J_{2}\left(z_{1j}\right)\right]^{2}}{2}\left[\frac{z_{1j}^{2}}{2\mu R^{2}}-E_{0}\pm\omega\right]+\mathcal{I}_{jj}\right\}+ \nonumber \\
& +\sum_{n\neq j}^{N}c_{n}^{\pm}\mathcal{I}_{jn} =-\mathcal{S}_{j},\label{eq:eq_min}
\end{align}
with $j\in\left\{ 1,2,...,N\right\} $. This equation defines a linear
system of equations whose solution determines the values of the coefficients
$c_{n}^{\pm}$. 

We can write Eq. (\ref{eq:eq_min}) in a more concise manner, using
matrix notation, as
\begin{equation}
\mathbb{M}\cdot\mathbf{c}^{\pm}=-\mathbf{S},
\end{equation}
where $\mathbf{c}^{\pm}$ and $\mathbf{S}$ are column vectors defined
as
\begin{align}
\left[\mathbf{c}^{\pm}\right]^{{\rm T}} & =\left(c_{1}^{\pm},c_{2}^{\pm},\ldots,c_{N}^{\pm}\right), \nonumber \\ \mathbf{S}^{{\rm T}} & =\left(\mathcal{S}_{1},\mathcal{S}_{2},\ldots,\mathcal{S}_{N}\right),\label{eq:M_matrix}
\end{align}
and $\mathbb{M}$ is a $N\times N$ matrix with:
\begin{equation}
\left(\mathbb{M}\right)_{ij}=g_{i}^{\pm}\left(\omega\right)\delta_{ij}+\mathcal{I}_{ij},
\end{equation}
where $\delta_{ij}$ is the Kronecker delta and 
\begin{equation}
g_{i}^{\pm}\left(\omega\right)=\frac{R^{2}\left[J_{2}\left(z_{1i}\right)\right]^{2}}{2}\left[\frac{z_{1i}^{2}}{2\mu R^{2}}-E_{0}\pm\omega\right].\label{eq:g_vec}
\end{equation}
After $\mathbb{M}$ and $\mathbf{S}$ are computed using Eqs. (\ref{eq: I integral}) and (\ref{eq:g_vec}),
and (\ref{eq:S integral}), respectively, the coefficients that determine
$\psi_{1}^{\pm}\left(\mathbf{r}\right)$ are readily obtained as
\begin{equation}
\mathbf{c}^{\pm}=-\mathbb{M}^{-1}\cdot\mathbf{S},\label{eq:c_coeffs}
\end{equation}
and the solution of the differential equation is found. 

\section{Linear Polarizability for interlayer and intralayer excitons \label{app:B}}
In this appendix we will make a quick comparison between the linear polarizability of both interlayer and intralayer excitons in a $\mathrm{WSe}_{2}/\mathrm{WS}_{2}$ heterobilayer. 

Intralayer excitons are considered for both the $\mathrm{WSe}_{2}$ and the $\mathrm{WS}_{2}$ layers, with the relevant material-dependent parameters. Additionally, due to the Coulomb-like  divergence of the potential at $r=0$, we consider the more appropriate variational \textit{ansatz}, defined by Pedersen \cite{PhysRevB.94.125424}, given by 
\begin{equation}
\psi_{0}\left(r\right)=\mathcal{C}\left(e^{-a r}-be^{- \gamma a r}\right)\left(R-r\right).\label{eq:exponential-ansatz}
\end{equation}

Minimizing this functional, the ground--state energy for excitons each layer will be 
\begin{align}
E_{\mathrm{WS}_{2}} = -0.00438405 = -119.296\;\mathrm{meV},\nonumber\\
E_{\mathrm{WSe}_{2}} = -0.00541680 = -147.399\;\mathrm{meV},\label{eq:monolayer_energies}
\end{align}
both closely matching the energies obtained via the finite basis approach from Sec.\ref{sec:besselJ-basis}. These binding energies are also substantially larger than the ground--state energy for interlayer excitons in the heterobilayer (Table \ref{tab:Comparison-of-the-1}), as expected from experimental studies \cite{Merkl2019}.

\begin{figure}
\centering
\begin{centering}
\includegraphics[scale=0.6]{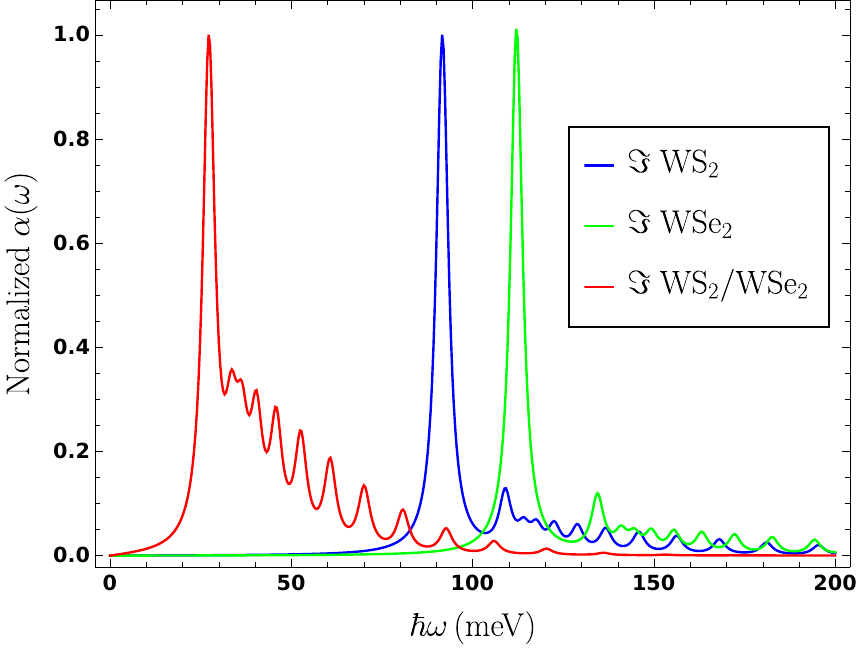}
\par\end{centering}
\caption{Imaginary part of the normalized linear polarizability for interlayer excitons in a $\mathrm{WSe}_{2}/\mathrm{WS}_{2}$ heterostructure and intralayer excitons in both a $\mathrm{WS}_{2}$ and a $\mathrm{WSe}_{2}$ monolayers. The peaks correspond to transitions between the $1s$ state (ground--state) and the excited $np$ states.}\label{fig:inter-intra_comparison}
\end{figure}

Computing the linear polarizability with the formalism described in Appendix \ref{app:A}, normalized by the value at the $1s \rightarrow 2p$ peak, we display the results in Fig. \ref{fig:inter-intra_comparison}.

\section{Variational Coefficients for the Third--Order Polarizability\label{app:C}}
Defining $g_{n}^{\left(l\right)}\left(\omega\right)$ as a more general form of Eq. (\ref{eq:g_vec})
\[
g_{n}^{\left(l\right)}\left(\omega\right)=\frac{R^{2}\left[J_{\left|l\right|+1}\left(z_{l,n}\right)\right]^{2}}{2}\left[\frac{z_{l,n}^{2}}{2\mu R^{2}}-E_{0}+\omega\right]
\]
and $\mathcal{V}_{nm}^{\left(l\right)}$ analogously to Eq. (\ref{eq: I integral})
\begin{align*}
\mathcal{V}_{nm}^{\left(l\right)} & =\left\langle J_{l}\left(z_{l,n}\frac{r}{R}\right)\middle| V_{i-RK} \left(r\right)\middle|J_{l}\left(z_{l,m}\frac{r}{R}\right)\right\rangle \\
& = \int_{0}^{R}J_{l}\left(z_{l,n}\frac{r}{R}\right) V_{i-RK} J_{l}\left(z_{l,m}\frac{r}{R}\right) rdr,
\end{align*} 
the functional of Eq. (\ref{eq:functionalK}) can then be rewritten using the both the $c^{\pm}_{n}$ and the $\zeta_{l,n}$ coefficients as
\begin{align}
\mathcal{K} & = \sum_{n=1}^{+\infty}\sum_{\left|l\right|=0}^{+\infty}\left[\zeta_{l,n}\right]^{\dagger}g_{n}^{\left(l\right)}\left(\omega_{a}+\omega_{b}\right)\zeta_{l,n}+\nonumber \\
& \quad+\sum_{n,m=1}^{+\infty}\sum_{\left|l\right|=0}^{+\infty}\left[\zeta_{l,n}\right]^{\dagger}\mathcal{V}_{nm}^{\left(l\right)}\zeta_{l,m}+\nonumber \\
& \quad + \frac{1}{2}\sum_{n,m=1}^{+\infty}\left\lbrace\left[\zeta_{0,n}\right]^{\dagger}\mathcal{T}_{nm}^{l=0}\cdot\left(c_{m}^{+}+c_{m}^{-}\right) +\right. \nonumber \\
&\left. + \left[\zeta_{\left|l\right|=2,n}\right]^{\dagger}\left(\mathcal{T}_{nm}^{l=2}\cdot c_{m}^{+}+\mathcal{T}_{nm}^{l=-2}\cdot c_{m}^{-}\right)+c.c.\right\rbrace. \label{eq:functional_expand}
\end{align}

In Eq. (\ref{eq:functional_expand}), $\mathcal{T}_{nm}^{l}$ is the dipole transition amplitude between  the functions of the basis, denoted by $\left\langle \psi_{\alpha}\left(\omega_{a}\right)\middle|\mathbf{d}_{\beta}\middle|\xi_{\alpha\beta}\left(\omega_{a},\omega_{b}\right)\right\rangle$ in Eq. (\ref{eq:functionalK}) and given by 
\begin{align*}
\mathcal{T}_{nm}^{l}& =\left\langle J_{l}\left(z_{l,n}\frac{r}{R}\right)\middle| \,r\, \middle|J_{1}\left(z_{1,m}\frac{r}{R}\right)\right\rangle \\
& = \int_{0}^{R}J_{l}\left(z_{l,n}\frac{r}{R}\right) r J_{1}\left(z_{1,m}\frac{r}{R}\right) rdr.
\end{align*}
The solution of this integral can be written in a closed form by applying the more general expression 
\begin{align}
& \int_{0}^{1}J_{\nu}\left(\alpha r\right) r J_{\nu+1}\left(\beta r\right) rdr = \frac{\alpha J_{\nu+1}\left(\alpha\right)}{\left(\alpha^{2}-\beta^{2}\right)^{2}}\times \nonumber \\  
& \qquad\times\left[-2\beta J_{\nu}\left(\beta\right)+\left(\alpha^{2}-\beta^{2}\right)J_{\nu+1}\left(\beta\right)\right], \label{eq:transfer_integral}
\end{align}
valid as long as $J_{\nu}\left(\alpha\right)=0$. Comparing with the definition of $\mathcal{T}_{nm}^{l}$, Eq. (\ref{eq:transfer_integral}) can be simplified further as $J_{\nu+1}\left(\beta\right)=0$. Performing the necessary changes of variable, and substituting $J_{\nu+1}\left(\beta\right)$ into Eq. (\ref{eq:transfer_integral}), $\mathcal{T}_{nm}^{l}$ can be written as 
\begin{align}
\mathcal{T}_{nm}^{\left(l = 0\right)} &= 
-2R^{3}\frac{z_{0,n}z_{1,m}J_{1}\left(z_{0,n}\right)J_{0}\left(z_{1,m}\right)}{\left(z_{0,n}^{2}-z_{1,m}^{2}\right)^{2}}, \nonumber \\
\mathcal{T}_{nm}^{\left(l = 2\right)} &= 
-2R^{3}\frac{z_{1,n}z_{2,m}J_{2}\left(z_{1,n}\right)J_{1}\left(z_{2,m}\right)}{\left(z_{1,n}^{2}-z_{2,m}^{2}\right)^{2}}.\label{eq:T_matrix}
\end{align}

As done in the appendix of \cite{henriques2021calculation}, we separate the $l=0$ and the $l=2$ cases. Looking first at the more complicated case of $l=0$, we must ensure that $\left|\xi_{\alpha\beta}\right\rangle $ is orthogonal to the considered ground--state [(in our case, the \textit{ansatz} of Eq. (\ref{eq:gaussian_ansatz-1})]. As such, to ensure this, we must have 
\begin{align}
\left\langle\psi_{0}\middle|\xi_{\alpha\beta}\right\rangle & = \sum_{n=1}^{+\infty}\zeta_{0,n}^{\alpha,\beta}\left(\omega_{a},\omega_{b}\right)\left\langle\psi_{0}\middle|J_{0}\left(z_{0,n}\frac{r}{R}\right)\right\rangle \nonumber \\
& = 0,
\end{align}
where non--zero angular momentum terms vanish upon angular integration due to the isotropic nature of the ground--state \textit{ansatz}. Separating the first term of the sum, this condition is given by 
\begin{align*}
&\zeta_{0,1}^{\alpha,\beta}\left(\omega_{a},\omega_{b}\right)\left\langle\psi_{0}\middle|J_{0}\left(z_{0,1}\frac{r}{R}\right)\right\rangle + \\
&+ \sum_{n=2}^{+\infty}\zeta_{0,n}^{\alpha,\beta}\left(\omega_{a},\omega_{b}\right)\left\langle\psi_{0}\middle|J_{0}\left(z_{0,n}\frac{r}{R}\right)\right\rangle = 0,
\end{align*}
which can then be rewritten as 
\begin{align}
&\zeta_{0,1}^{\alpha,\beta}\left(\omega_{a},\omega_{b}\right) =  \nonumber \\
&=-\sum_{n=2}^{+\infty}\zeta_{0,n}^{\alpha,\beta}\left(\omega_{a},\omega_{b}\right)\frac{\left\langle\psi_{0}\middle|J_{0}\left(z_{0,n}\frac{r}{R}\right)\right\rangle}{\left\langle\psi_{0}\middle|J_{0}\left(z_{0,1}\frac{r}{R}\right)\right\rangle}.\label{eq:orthogonality}
\end{align}
This means that $\zeta_{0,1}$ is no longer considered an independent variable and, as such, we will focus our attention on the $n\geq2$ terms. For compactness, we define
\begin{align}
\mathbf{f}_{n} = \frac{\left\langle\psi_{0}\middle|J_{0}\left(z_{0,n}\frac{r}{R}\right)\right\rangle}{\left\langle\psi_{0}\middle|J_{0}\left(z_{0,1}\frac{r}{R}\right)\right\rangle},
\end{align}
which lets us rewrite Eq. (\ref{eq:orthogonality}) as
\begin{align}
\zeta_{0,1}^{\alpha,\beta}\left(\omega_{a},\omega_{b}\right) = -\sum_{n=2}^{+\infty}\zeta_{0,n}^{\alpha,\beta}\left(\omega_{a},\omega_{b}\right)\mathbf{f}_{n}. \label{eq:zeta01}
\end{align}

Substituting Eq. (\ref{eq:zeta01}) into Eq. (\ref{eq:functional_expand}) and then differentiating the resulting expression with respect to $\left[\zeta_{l,n}\right]^{\dagger}$, we obtain for $n\geq2$, 
\begin{widetext}
\begin{align}
\left[g_{1}^{\left(0\right)}\left(\omega_{a}+\omega_{b}\right)\mathbf{f}_{n}\mathbf{f}_{m}+\mathcal{V}_{1,1}^{\left(0\right)}\mathbf{f}_{n}\mathbf{f}_{m}\right]\zeta_{0,n}-\mathcal{V}_{1,m}^{\left(0\right)}\mathbf{f}_{n}\zeta_{0,n}-\mathcal{V}_{n,1}^{\left(0\right)}\mathbf{f}_{m}\zeta_{0,n} - \frac{1}{2}\mathbf{f}_{n}\sum_{m=1}^{+\infty}\mathcal{T}_{nm}^{l=0}\cdot\left(c_{m}^{+}+c_{m}^{-}\right) + & \nonumber \\ 
+\sum_{n=2}^{+\infty}g_{n}^{\left(0\right)}\left(\omega_{a}+\omega_{b}\right)\zeta_{0,n}+\sum_{n,m=2}^{+\infty}\mathcal{V}_{nm}^{\left(0\right)}\zeta_{0,m} + \frac{1}{2}\sum_{n,m=2}^{+\infty}\left\lbrace\mathcal{T}_{nm}^{l=0}\cdot\left(c_{m}^{+}+c_{m}^{-}\right) +c.c.\right\rbrace & =0.
\end{align}
This means that the system can be easily solved algebraically as 
\begin{align}
\mathbf{\zeta}_{0}^{x,x}\left(\omega_{a},\omega_{b}\right)=\left[\mathbb{F}+\mathbb{M}^{\left(0\right)}\left(\omega_{a}+\omega_{b}\right)\right]^{-1}\cdot\left[\mathbf{W}_{0}^{x,x}\left(\omega_{a}\right)+\mathbf{f}_{0}^{x,x}\left(\omega_{a}\right)\right],\label{eq:zeta_l0}
\end{align}
\end{widetext}
with $\mathbb{M}^{\left(0\right)}\left(\omega_{a}+\omega_{b}\right)$ defined as in Eq. (\ref{eq:M_matrix}) and
\begin{equation*}
\mathbf{W}_{0}^{x,x} = - \mathbb{T}^{0}\cdot\mathbf{c} ,
\end{equation*}
\begin{equation}
\left(\mathbf{f}_{0}^{x,x}\right)_{n} = \mathbf{f}_{n}\sum_{m=1}^{N}c_{m}\mathcal{T}_{1,m}^{\left(0\right)}c_{m},
\end{equation}
\begin{align*}
\left(\mathbb{F}\right)_{i,j} & = \left[g_{1}^{\left(0\right)}\left(\omega_{a}+\omega_{b}\right)+\mathcal{V}_{1,1}^{\left(0\right)}\right]\mathbf{f}_{i}\mathbf{f}_{j}-\nonumber \\ & \qquad -\mathcal{V}_{1,j}^{\left(0\right)}\mathbf{f}_{i}-\mathcal{V}_{i,1}^{\left(0\right)}\mathbf{f}_{j}.
\end{align*}
The missing coefficient $\zeta_{0,1}^{\alpha,\beta}\left(\omega_{a},\omega_{b}\right)$ is obtained by calculating Eq. (\ref{eq:zeta01}) with the solution of Eq. (\ref{eq:zeta_l0}).
Several expressions have been already simplified by taking advantage of the parity of Bessel functions of the first kind, 
\begin{equation*}
J_{-l}\left(x\right) =(-1)^lJ_{l}\left(x\right).
\end{equation*}

Looking now at the simpler $l=2$ case, and as no orthogonality--based restrictions need to be applied to its coefficients, we can directly differentiate Eq. (\ref{eq:functional_expand}) with respect to $\left[\zeta_{l,n}\right]^{\dagger}$ and then minimize the resulting expression, obtaining 
\begin{align}
\sum_{n=1}^{+\infty}g_{n}^{\left(2\right)}\left(\omega_{a}+\omega_{b}\right)\zeta_{2,n}+\sum_{n,m=1}^{+\infty}\mathcal{V}_{nm}^{\left(2\right)}\zeta_{2,m}+ & \nonumber \\
+ \frac{1}{2}\sum_{n,m=1}^{+\infty}\left\lbrace\mathcal{T}_{nm}^{l=2}\cdot c_{m}^{+}+\mathcal{T}_{nm}^{l=-2}\cdot c_{m}^{-}\right\rbrace =&0.
\end{align}
The $\zeta_{2,n}$ variational coefficients can then be obtained, in vector form, as 
\begin{align}
\mathbf{\zeta}_{2}^{x,x}\left(\omega_{a},\omega_{b}\right)=\left[\mathbb{M}^{\left(2\right)}\left(\omega_{a}+\omega_{b}\right)\right]^{-1}\cdot\mathbf{W}_{2}^{x,x}\left(\omega_{a}\right),\label{eq:zeta_l2}
\end{align}
a similar system to the one in Eq. (\ref{eq:c_coeffs}) with $\mathbf{W}_{2}^{x,x}$ defined analogously to $\mathbf{W}_{0}^{x,x}$, after simplifications from parity of Bessel functions.

\bibliographystyle{unsrt}
\bibliography{interlayer.bib}

\end{document}